\documentclass[aps,prl,twocolumn,superscriptaddress,showpacs]{revtex4}
\usepackage{graphicx}
\begin{document}
\title{\bf Comment on ``Detecting the Kondo Screening Cloud around a Quantum
Dot''}

\author{A.~A.~Zvyagin}
\affiliation{B.~I.~Verkin Institute for Low Temperature Physics and
Engineering of the NAS of Ukraine, Kharkov, 61164, Ukraine}

\date{\today}

\begin{abstract}
We point out several mistakes in the recent work of I.~Affleck and
P.~Simon (cond-mat/0012002). 
\end{abstract}

\maketitle

Recently \cite{1} it was claimed that the effect of the Kondo screening on 
persistent currents (PC) was studied in a mesoscopic ring coupled to a 
quantum dot. The system was considered in the framework of the Anderson 
impurity (AI) model for two situations --- for large Kondo length scales 
$\xi_K \gg L$ and small ones $\xi_K \ll L$. Here $\xi_K = \hbar v_F /T_K$, 
$v_F$ is the Fermi velocity of electrons in the ring and $T_K$ is the Kondo 
energy scale. We want to point out the following: \\
(i) Charge degrees of freedom (and, hence, PC) are {\em totally decoupled} 
from spin degrees of freedom in the ``bare'' Kondo problem \cite{2}. Hence, 
the Kondo screening itself {\em does not} affect (except of initial phase 
shifts) charge PC (the Aharonov-Bohm effect), but rather spin PC (the 
Aharonov-Casher one) \cite{3}. On the other hand, the AI (for which both spin 
{\em and charge} degrees of freedom are hybridized with the host) really 
influences charge PC \cite{4}. \\
(ii) In the case $\xi_K \gg L$ (i.e., $T_K$ being much less than energy 
spacings of electrons in the ring $\hbar v_F/L$) one can neglect all other 
levels of electrons in the ring except of the lowest. (Sinusoidal 
oscillations of PC with the magnetic flux obviously results.) In this case 
the renormalization of the ground state energy of the localized electron due 
to the hybridization with ones from the ring (i.e., $T_K$ \cite{2}) is {\em 
not} determined by Eq.~(1) of \cite{1}. In the limit $|\epsilon_0|, U \gg t'$ 
\cite{1} it is rather proportional to $2t'^2/|(\hbar v_F/L) + 
[(U+\epsilon_0)\epsilon_0/U]|$, where $t'$ are hopping elements of dot-ring 
contacts, $U$ and $\epsilon_0$ are the standard parameters of the AI 
Hamiltonian \cite{1,2}. Kondo logarithms (which appear if one takes into 
account {\em thermodynamically large number} of states in the Fermi sea of 
ring electrons) are absent in this case \cite{2}. Here one can speak about the 
Kondo effect only in a ``Pickwick sense'' (the Abrikosov-Suhl resonance is 
too far from the Fermi energy and too wide). The other situation $\xi_K \ll 
L$ ($T_K \gg \hbar v_F/L$) was studied {\em exactly} 6 years ago in 
\cite{3,4} (unlike the {\em qualitative} study of this case in \cite{1}). 
Here a thermodynamically large number of states of electrons of the ring 
produces exponential, generic for the {\em correlated nature} of the Kondo 
effect, dependence for $T_K$ \cite{2}. Hence, in the ground state the 
oscillations of PC are ``saw-tooth''-like (the sum of many harmonics) 
\cite{3,4,5}. Their magnitude is proportional to $ev_F/\hbar L$ \cite{4,5}. 
However the conclusion of \cite{1} that PC in this case are those of {\em 
ideal} ring (i.e., without dot) is invalid. The intra-dot Coulomb repulsion 
manifests itself in the matrix of ``dressed charges'' (its components measure 
numbers of electrons which form low-lying charge and spin excitations of a 
system), being {\em different} from unity matrix (noninteracting case) 
\cite{3,4}. Only in the limit of {\em zero} magnetization the behavior of PC 
is reminiscent of the one for free electrons, while for any nonzero Zeeman 
splitting (which is the generic situation) it differs drastically. Moreover, 
the single velocity ($v_F$) present in the answer for PC is the consequence 
of the linearization of the dispersion law for electrons in the ring in the 
AI model. It is easy to show that the study of PC in a {\em lattice} Bethe 
ansatz-solvable model (e.g., \cite{6} where the AI-like impurity was situated 
in the correlated electron ring) produces the answer for the PC similar to 
the ones for the AI model \cite{4}, but with {\em two different velocities} 
for low-lying spin and charge excitations. The latters become equal to $v_F$ 
only if one linearizes the spectrum of the host. \\
(iii) PC for the {\em multichannel} Kondo impurity were {\em exactly} studied 
in \cite{7} (notice that the results of \cite{7} can be used for chiral 
fermions, too). Here multichannel Kondo screening also {\em does not} affect 
charge PC (as the consequence of the spin-charge separation) for {\em any}
values of the local exchanges between the localized spin and channel electrons
of the ring. However the correlation effects induced by the magnetic impurity 
introduce the interference of several Aharonov-Casher oscillation patterns 
for spin PC. There is also an interference between the Coulomb blockade-like 
(parity) effects and PC oscillations \cite{7}. 

This has been the text of my manuscript submitted to the Physical Review
Letters on March 26, 2001. Several remarks are in order to clarify the
brief above argumentation.   

1. A number of recent publications considered persistent currents in 
quantum rings with magnetic impurities. There is a principal
difference between persistent currents and usual transport currents. 
Recall, usual currents are {\em transport}, kinetic characteristics of
any system. Their usual characteristics are the resistivity or
conductivity and related to them transition amplitude. In the
framework of the linear response theory the conductivity is the
coefficient, which connects the value of the current with the value of
an applied {\em electric} field. Hence, one can consider transport
currents in a system only if it has the sourse and drain,
cf. Fig.~1~(a), and the transport current is the consequence of the 
difference in potentials applied to the sourse and drain. One
necessary needs this difference to study any transport current.  
\begin{figure}
\begin{center}
\includegraphics[height=6cm,width=0.4\textwidth]{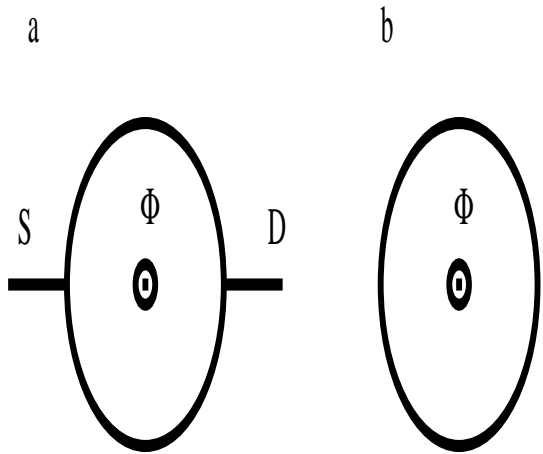}
\end{center}
\caption{\label{fig1} Different geometries for the manifestation of
the Aharonov-Bohm effect of an external magnetic flux $\Phi$ in a
metallic ring: (a) the transport current geometry with the source (S)
and drain (D); (b) the persistent current geometry.}
\end{figure}
On the other hand, the persistent current is the {\em thermodynamic} 
characteristic of a ring. It is connected with the Aharonov-Bohm phase
shift, which appears when charges move along a loop, pierced by a magnetic
flux \cite{ABC}. Then an external magnetic flux yields nonzero
momentum of charges. In fact the persistent current is nothing else
than the {\em total orbital moment} of all charges in the
ring. Naturally, the persistent current can exist without any applied 
external electric field; it does not need any sourse and drain,
cf. Fig.~1~(b). Persistent current is just the derivative of the
energy of a system in equilibrium with respect to the applied magnetic
flux. For noninteracting electron rings the charge persistent
current is connected with the virtual movement of an electron around the
magnetic flux. For interacting electron systems charge persistent
currents pertain to the virtual movement(s) of charge-carrying
excitations (in the ground state and at low temperatures --- to
the virtual movement of the low-lying charged excitations). Clearly,
in the situation of Fig.~1~(b) (without any sourse and drain) it is 
inappropriate to speak about the conductance/conductivity or
resistivity of a system --- at least one should first properly define,
what are the latters. The missunderstanding sometimes appears because 
several recent experiments on quantum rings studied namely the
geometry of Fig.~1~(a), but not of Fig.~1~(b) (i.e., they measured
transport currents between sourses and drains, but not the total
orbital moment of the ring). Naturally, when the ring between the
sourse and drain is pierced by an external magnetic flux in the
geometry of Fig.~1~(a), the transport current is also affected by that
flux. Hence, the conductance of the transport current also becomes
flux-dependent. However this transport current is naturally {\em not}
exactly equal to the persistent current.  It turns out that the
authors of Ref.~\cite{1} definitly wrote about persistent currents,
not transport currents (cf., Fig.~1 and Fig.~1 of \cite{1}).  

This difference in the basic features of transport and persistent 
currents produces the main difference in the answers, when one
considers the effect of a magnetic (Kondo) impurity. The Kondo
impurity introduces the interaction into the free electron problem
\cite{2}. It is well known that in the pure Kondo problem (a magnetic 
impurity, just a spin, coupled to the free electron host via a local 
exchange interaction) collective spin and charge degrees of freedom 
decouple from each other \cite{2}. Notice that in the Anderson
impurity model the hybridization impurity manifests both charge and
spin degrees of freedom \cite{2}. On the one hand, the resistivity of
a transport current is strongly affected by the Kondo impurity
\cite{2}, because the resistivity is determined by the density of {\em
all} states of the system, spin and charge. The Kondo 
(Abrikosov-Suhl) resonance (which is the characteristic feature of
{\em spin} degrees of freedom, cf., \cite{2,KO}) determines the 
magnetoresistivity of the transport current. Hence, the conductance of
a quantum ring [in the geometry of Fig.~1~(a)] with embedded magnetic
(Kondo) impurity is, obviously, affected by that impurity. On the
other hand, the persistent current [i.e., the total orbital moment of
a quantum ring in the geometry of Fig.~1~(b)] is determined by the
virtual movement of only {\em charge-carrying excitations} around the 
magnetic flux. In the pure Kondo situation, in which a magnetic
impurity (spin) is connected to the free electron host via a local
spin exchange, the charge degrees of freedom are mostly {\em not}
affected by the spin impurity \cite{2} (the only possible effect is
the possible presence of an initial phase shift, see 4. below). Hence,
in this case the Kondo impurity does not influence the frequency and
the magnitude of persistent currents in the geometry of
Fig.~1~(b). Namely this property of metallic rings with embedded spin
(Kondo) impurity was pointed out in our pioneering Letter of 1994
\cite{3}, which studied the effect of the Kondo impurity on persistent
currents for the first time. The situation is very diffrent for the
Anderson (hybridization) impurity. Here the impurity affects both spin
and charge low-lying excitations \cite{2}. In this case the Anderson
impurity affects charge persistent currents of the geometry of
Fig.~1~(b) \cite{4}. For example, the Anderson impurity produces
low-lying excitations, which carry charge $-2e$ (spin-singlet bound
states of electrons) and excitations, which carry spin $1/2$ and
charge $-e$ \cite{2}. The virtual movement of excitations, which carry
charge $-2e$, naturally yields oscillations of charge persistent
currents with the period $\Phi_0/2$ (where $\Phi_0 = hc/e$), while
unbound electron excitations, which carry charge $-e$, produce
oscillations of persistent currents with the period $\Phi_0$
\cite{4}. For the case with the dispersion law of host electrons being
linearized about Fermi points the velocities of both types of
low-lying excitations are equal to each other (and both are equal to
the Fermi velocity) \cite{2}. In the absence of the Zeeman effect of
the external magnetic field the interference of those two type of 
oscillations of persistent currents produces oscillations with the
period $\Phi_0$, reminiscent of the ones in a free electron
host. (Notice also the parity effect of persistent currents in a ring
with the magnetic impuritry, i.e., different initial phases
[dia- or paramagnetic persistent currents] and periodicities for
different numbers of electrons in the ring, predicted in
\cite{3,4,5,7}). However the nonzero curvature of the spectrum of host
electrons and the Zeeman effect of the applied magnetic field (it is
small, naturally, for the case of GaAs-based quantum rings, where
effective $g$-factors are small) must produce the real interfererence
of two types of oscillations of charge persistent currents (with the
periods $\Phi_0$ and $\Phi_0/2$), because their magnitudes become
different from each other in that case. It turns out that we pointed
out the difference between the characteristics of persistent currents
and transport currents of a metallic ring with a Kondo impurity in 
Ref.~\cite{7}, where it is clearly shown that a magnetic Kondo
impurity (spin) does not change the properties of charge persistent
currents, but drastically affects the magnetoresistivity for charge
transport currents, cf. subsections A and B of the Section~IV of \cite{7}.   

2. To explain other inconsistencies of \cite{1} let us start with the 
description of the low-energy behavior of the Anderson model. In this 
explanation we use different from \cite{2}, variational approach to
the Kondo problem \cite{var}. This approach is well-known and later
was reproduced in a number of books and textbooks, see e.g., 
\cite{Ful,Mah}. Naturally similar results can be obtained using the
Bethe ansatz approach to the Kondo problem \cite{2}, or using other
approaches, like the non-crossing-like approximations \cite{BCW} or
the slave-boson technique \cite{Col}. For simplicity we consider 
the Anderson impurity Hamiltonian with $U \to \infty$ (similar results
can be obtained for finite $U$). The variational wavefunction for the
ground state can be taken as 
\begin{equation}
|\psi \rangle = A(|0 \rangle  + \sum_{\epsilon} a(\epsilon)|\epsilon
\rangle ) \ , 
\end{equation}
where $|0 \rangle$ determines the state of host electrons, in which
all states below the Fermi energy are occupied and the impurity level
is empty, $|\epsilon \rangle$ is the state with one electron at the
impurity level and one hole below the Fermi energy, $a(\epsilon)$ is
the variational coefficient, and $A$ is the normalization
constant. The minimization of the ground state energy with respect to
$a$ yields the equation 
\begin{equation}
\Delta E = t'^2 \sum_{\epsilon} {2 \over (\Delta E - \epsilon_0 + 
\epsilon)} \ , 
\end{equation}
where we introduced $\Delta E$, the renormalization of the ground
state energy due to the Anderson impurity. If the number of states in
the sum is thermodynamically large (namely it is the case for the
small ratio $\hbar v_F /L$), one can replace the sum by the integral,
which for the constant density of states of conduction electrons
$N(0)$ yields the famous Kondo logarithm and one obtains: 
\begin{equation}
\Delta E = - De^{-{|\epsilon_0|\over 2N(0)t'^2}} = 
-De^{-{1\over N(0)J_{eff}}} \equiv -T_K 
\end{equation}
($D$ pertains to the bandwidth of conduction electrons). It means that
$T_K$ is namely the renormalization of the ground state energy due to
the hybridization Anderson impurity. This, naturally, agrees with
other definitions of the Kondo scale, see, e.g., Refs.~\cite{2}. This 
equation is equivalent, naturally, to Eq.~(1) of \cite{1}. In this
case, obviously, the thermodynamically large number of harmonics
produces the ``saw-tooth''-like oscillations of persistent currents,
which were obtained in Refs.~\cite{3,4,5,7} and (partially) reproduced
in \cite{1}. However, it turns out that if the number of states in the
sum is small (which is the case for the large ratio $\hbar v_F/L$),
one {\em cannot} replace the sum by the integral, and the Kondo
logarithm does not appear in Eq.~(2). One can, naturally, find the 
renormalization of the ground state energy (which also plays the role
of the Kondo temperature in this case), however it will be {\em not} 
exponentially dependent on the effective exchange constant
$J_{eff}$. For one electron level of the ring being involved it is
namely the answer written above in the text of the Comment (in the
limit $U \to \infty$; for finite $U$ one can follow for the
variational approach \cite{GS}). This is actually the answer obtained
in \cite{BS}, to which the limiting case of the large ratio $\hbar
v_F/L$ of \cite{1} pertains (\cite{1} actually introduces only the $O(J^2)$
corrections to the result \cite{BS}). Naturally, 
for only one level of conduction electrons being involved one clearly
obtains only one harmonics (i.e., sinusoidal oscillations of the
persistent current, cf. \cite{1,BS}). This means that in the limiting
case of large $\hbar v_F/L$ one {\em cannot} use Eq.~(1) of \cite{1}
for the determination of $T_K$. Actually this limit ($\xi_K \gg L$)
has no other, standard for the Kondo situation, features. For example,
it is well known that the generic Kondo effect affects the Sommerfeld
coefficient of the low temperature specific heat and magnetic
susceptibility of the magnetic impurity: They become large (inverse
proportional to the small Kondo temperature). However for the finite
number of states of the case $\xi_K \gg L$ the specific heat is
exponentially small at low temperatures, and the magnetic
susceptibility is determined by the (possible for systems with orbital
degrees of freedom) van Vleck (zero temperature) terms, while
temperature corrections are exponentially small, too \cite{Sch}. 

In Ref.~\cite{4} persistent currents were calculated for the ring with
the Anderson impurity (which produces effective interactions for {\em
both} spin and charge degrees of freedom). In this paper it is shown
that persistent currents are determined by $1/L^2$ corrections to the
energy ($L$ is the size of the ring), while the behavior of the
impurity itself (its valence, magnetization, etc.) are determined by 
$1/L$ corrections, if the main contribution to the energy is of order
of 1 (i.e., per site). [This statement is true for {\em any}
one-dimensional quantum ring with gapless low-lying excitations.] The
Kondo scale in the magnetic field behavior appears for the
characteristics of the impurity, but not (in the main order in $1/L$)
for persistent currents (it appears in higher-order corrections, too,
but they are irrelevant for our discussion). The main quantities,
which determine the values of persistent currents in ideal ballistic
quantum mesoscopic rings are the (Fermi) velocities of low-lying
excitations, which virtual movement defines the Aharonov-Bohm effect in 
interacting systems, and the matrix of ``dressed charges'' (this quantity 
measures the effective number of ``initial'' electrons which form each
low-lying excitation). In principle both of those quantities depend on
the magnetic field (in the Bethe ansatz scheme that dependence is
obtained via solutions of integral equations for mentioned
quantities). For the model of the Anderson impurity studied in
Ref.~\cite{4} (with the linearized dispersion law of host electrons)
the velocities of low-lying charge and spin excitations coincided with
the initial Fermi velocity of electrons and do not depend on the
external magnetic field (cf. \cite{2}). Notice that for 
hybridization impurity models on the lattice, cf. \cite{6}, it is not
true --- there are {\em two} different from each other velocities of 
low-lying excitations and both of them depend on the magnetic
field. However the matrix of ``dressed charges'' {\em does} depend on
the magnetic field even for the linearized dispersion law of host
electrons. This dependence does not reveal the Kondo scale (in the
main order in $1/L$). For {\em exactly} zero magnetic field the answer
for the charge persistent current is reminiscent of the one for a ring
of noninteracting electrons (as it must be). However even a small
deviation of the value of the field from zero produces very strong
changes in the values of the components of the ``dressed charge''
matrix [because of logarithmic corrections, present in the
SU(2)-symmetric system; these corrections are well-known to one of the
authors of \cite{1}, who published many papers devoted to similar 
logarithmic corrections \cite{log}]. Hence the persistent current
becomes to be different from the one of the ring with free electrons
for small $\hbar v_F/L$ (even if the value of the magnetic field is
small compared to $T_K$, and the magnetization of the impurity is
small). These corrections, being very small, are not very important
for the magnetization of the impurity, however they are of great
importance for persistent currents ---  e.g., the magnitudes of 
the oscillations of persistent currents with periods $\Phi_0$ and
$\Phi_0/2$ will be different even for very small fields, i.e., an
additional period of persistent currents can be observed. We emphasize
that it is not rare that the effect of an (even small) magnetic field
on the quantum dot embedded into the quantum ring is important (cf. 
\cite{field}).  

By the way, in a recent preprint \cite{prepr} (which appeared after
the authors of \cite{1} became aware of my above written text of the
Comment) one of the authors of \cite{1} directly wrote (cf. P.7 of 
\cite{prepr}): ``On the other hand, when $\xi_K \gg L$ the Kondo
effect doesn't take place: the infrared divergence of the Kondo
coupling, $\lambda$, is cut off by the finite size of the ring.'' This
is in obvious contradiction to the statements of \cite{1}, where the
authors proposed to study the influence of the Kondo effect on
persistent currents in {\em both} cases --- $\xi_K \gg L$ and 
$\xi_K \ll L$ --- and obviously coincide with the above mentioned
statement of our Comment. 

3. The authors of Ref.~\cite{1} (and also some others, see, e.g.,
\cite{side}) claimed that the Bethe ansatz approaches of 
Refs.~\cite{3,4,5,7,EJS} considered {\em side-coupled} quantum
dots, see Fig.~2~(a). 
\begin{figure}
\begin{center}
\includegraphics[height=6cm,width=0.4\textwidth]{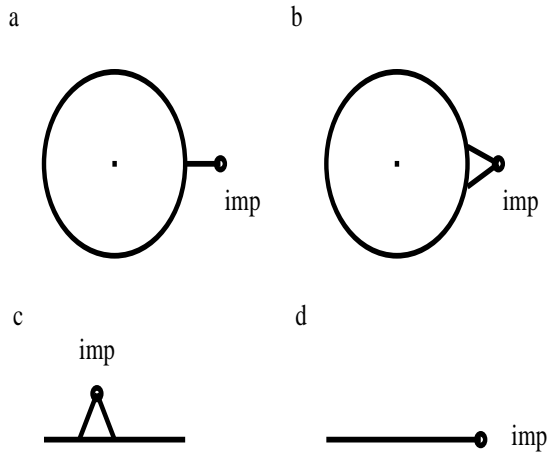}
\end{center}
\caption{\label{fig2} (a) A side-coupled impurity; (b) an integrable
impurity in a ring; (c) an integrable impurity in the bulk of an an
open chain; (d) an integrable impurity at the edge of an open chain.}
\end{figure}
These claims are absolutely wrong. The Bethe ansatz method can be used
either for systems with periodic boundary conditions, or for systems
with open boundary conditions (does not matter, whether one studies 
homogeneous systems or systems with impurities). The Bethe ansatz
method is properly justified for discrete coordinates of particles 
\cite{KBI}. Field theoretical models, solved by the Bethe ansatz, have
to be continuous limits of their lattice counterparts (including the
solution of the Kondo problem \cite{2}). (Notice that the continuous
limit can be taken in different ways, which introduces some
ambiguity.) Impurities can be included into the lattice Bethe ansatz
scheme either as shown in Fig.~2~(b) and (c) for the impurity in the
bulk of a ring or an open chain, i.e., the impurity is connected with 
{\em two} neighboring sites of the host, or connected with only one 
neigboring site only at the edge of an open chain, as shown in
Fig.~2~(d). This is the direct consequence of the fact that impurities
can be introduced into the Bethe ansatz monodromies either as special
{\em scattering} matrices (this way implies no reflection in the
problem at all) or as local {\em boundary fields/potentials}
(reflectors), which can be applied {\em only} to the edges of an open
chain. Please, pay attention that we distinguish the reflection and
backward scattering (by the latter we mean the transfer from the one
Fermi point to another --- such processes are present in any lattice 
integrable theories, like the Heisenberg spin-${1\over2}$ chain or
Hubbard chain \cite{KBI}, but there is no reflection there for
periodic boundary conditions). Scattering matrices of impurities have
to satisfy {\em Yang-Baxter} (``triangular'', ``star-triangle'') 
{\em relations} \cite{KBI} with scattering matrices of the host, to
preserve the integrability of a problem in the framework of the Bethe's
ansatz. Hense, such impurities in periodic Bethe ansatz solvable
systems have to be pure scatterers, but must {\em not} produce any 
reflection. On the other hand, local fields (reflectrors) can be used
only for {\em open} chains in the lattice Bethe ansatz approach. They
are described by reflection matrices, which satisfy {\em reflection 
equations} \cite{Skl}. However for any system with open boundary
conditions persistent currents are obviously zero. Hence, the only 
possibility to study persistent currents in Bethe ansatz-solvable
models with impurities is to consider impurities, which produce only 
scattering phases, but not reflections, and which scattering matrices
satisfy Yang-Baxter relations with scattering matrices of the host
(and, naturally, mutually). In our papers \cite{3,4,5,7} 
(see also \cite{oth}: It turns out that we studied the problem of the 
influence of magnetic impurities on charge and spin persistent
currents in detail in a large number of refereed papers since 1994 
and reported our results at international conferences), devoted to the
influence of magnetic and hybridization impurities on persistent
currents, we considered only integrable impurities of this class. The
Bethe ansatz solution of the Kondo problem \cite{2} also belongs to
this class --- impurities produces only scattering phases but not 
reflections. The side-coupled impurity [cf. Fig.~2~(a)] has,
naturally, the properties of a reflector (which is correctly pointed out
in \cite{1}) and, therefore, cannot be introduced into the Bethe
ansatz solvable ring in principle, because it violates the Yang-Baxter
relations. This is why, the claim that the side-coupled impurity could
be introduced into the Bethe ansatz solvable ring is absolutely
incorrect.     

4. In the framework of the Bethe ansatz there are two independent very
well known solutions to the Kondo problem (pure spin impurity in the free
electron host, coupled to the host via the local exchange): the one 
by N.~Andrei and the one by P.~Wiegmann \cite{2}. These solutions,
being different in some details, produce the same correct answers for 
thermodynamic characteristics of the Kondo magnetic impurity. Those
details are not important for the behavior of the
thermodynamic characteristics of the host and the impurity. However,
those details can produce the difference in the behavior of
finite-size corrections (which namely determine persistent currents in
metallic systems). One of those details is the phase factor for the
Bethe ansatz equations, which govern the behavior of charge degrees
of freedom in the solution of P.~Wiegmann (cf., \cite{Wie}) and the
absence of that phase in the approach of N.~Andrei (cf., \cite{And}),
see also \cite{2}. The presence or absence of those phase factors are 
determined by the particular choosen schemes of taking the sclaling
limit in those two approahes. The Bethe ansatz is developed for
systems with {\em discrete} particles, because of its main property: Any
multi-particle scattering process is considered as consequence of pair
scattering processes between particles in the Bethe ansatz scheme
\cite{KBI}. Then, when studying continuous limit, one has to use some
scaling approximation from the lattice counterpart. Therefore the 
consideration of the continuous scaling limit for the solution of the
Kondo problem has some freedom in the determination of the phase
shift. In  \cite{3} we used the scaling scheme intoduced by
P.~Wiegmann. This scheme determined the appearence of the initial
phase shift for charge persistent currents, caused by the Kondo
impurity. On the other hand, in Ref.~\cite{7} we studied persistent
currents for a system with multichannel Kondo impurities, and used
the scaling scheme introduced by N.~Andrei. This is why there was no 
initial phase shift for charge persistent currents (even for the
limiting case of the number of channels being equal to 1). However, as
we pointed out above, the presence or absence of that initial phase
shift is determined by the (non-controlable) scaling approximation. We
do not know, which answer (with or without the phase shift) is generic
in the real situation of a quantum dot in a ring. The only argument,
which can be used, is that for the {\em lattice} exactly solvable
problem of a magnetic impurity in a correlated electron ring
(cf. \cite{6}) the Bethe ansatz equations for charge degrees of
freedom {\em have} phase factors, which, naturally, produce initial
phase shifts for charge persistent currents. These initial phase
shifts are the consequences of the fact, that a magnetic impurity
introduces the nonzero net chirality into the periodic lattice
integrable problem.   

5. It turns out that our previous results \cite{3,7} for spin and
charge persistent currents in a metallic ring with a magnetic (spin)
Kondo impurity coincide with the ones of Ref.~\cite{EJS} and with the
limiting case of $\xi_K \ll L$ of \cite{1} for zero magnetization of
the system (up to the initial phase shift in \cite{3}, see the
discusssion in 4. above). In some of our studies we considered the
behavior of chiral (only right- or left-moving) electrons. Refs.~\cite{1,EJS}
consider both right- and left-movers together. However the fact that
the answers of Refs.~\cite{3,7} and \cite{1,EJS} are similar implies
that the chirality of host electrons does not play an important role
in this problem. By the way, naturally, for the linearized dispersion law
of host electrons one must consider for persistent currents only {\em 
finite-size corrections}, which depend on external fluxes. The
constant term (of order of the size of the system) is the obvious
artifact of the linearization of the spectra of host electrons and
was, obviously, discarded in Refs.~\cite{3,4,5,7,oth}.

\end{document}